\documentstyle[preprint,eqsecnum,aps]{revtex}
\begin{document}
\def\square{\kern1pt\vbox{\hrule height 1.2pt\hbox{\vrule width 1.2pt\hskip 3pt
   \vbox{\vskip 6pt}\hskip 3pt\vrule width 0.6pt}\hrule height 0.6pt}\kern1pt}

\def\lta{\mathrel{\spose{\lower 3pt\hbox{$\mathchar"218$}}
     \raise 2.0pt\hbox{$\mathchar"13C$}}}
\def\gta{\mathrel{\spose{\lower 3pt\hbox{$\mathchar"218$}}
     \raise 2.0pt\hbox{$\mathchar"13E$}}}
\def\spose#1{\hbox to 0pt{#1\hss}}

\draft
\preprint{CITA-94-34}
\title{Inflation from Extra Dimensions}
\author{Janna J. Levin }
\address{ Canadian Institute for Theoretical Astrophysics}
\address{Mc Lennan Labs, 60 St. George Street, Toronto, ON M5S 1A7}
\maketitle
\begin{abstract}

A gravity-driven inflation is shown to
arise from a simple higher dimensional universe.
In vacuum,
the shear of $n>1$ contracting dimensions is
able to  inflate the remaining three spatial dimensions.
Said another way, the expansion of the 3-volume
is accelerated by the contraction of the $n$-volume.
Upon dimensional reduction, the theory is
equivalent to a four dimensional cosmology
with a dynamical Planck mass.
A connection can therefore be made to recent
examples of inflation
powered by a dilaton kinetic energy.
Unfortunately, the graceful exit
problem encountered
in dilaton cosmologies will haunt this
cosmology as well.

\vskip10pt

98.80.Hw, 98.80.Cq,04.50.+h

\bigskip

\centerline{to be published in {\it Phys. Lett} B}

\end{abstract}

\narrowtext
\vfill\eject

\section{ }

Imagine a spacetime which is created empty.
Even in the absence of matter as a source of fuel,
the evolution of the spacetime can
still involve interesting dynamics.
The shearing
and twisting of dimensions can feed each other,
driving some dimensions to expand and others to
contract.

Suppose a universe is created initially
with some spatial dimensions contracting
and the others expanding.
The shear of the contracting dimensions
has been shown to expand the other dimensions at a humble
pace \cite{chodos}.
The expansion of the  spatial sections decelerates with time.
It is the intent of this paper to present an additional
possibility.
The contraction of some dimensions can actually inflate
the expanding space; that is, the expansion of the
spatial sections can accelerate with time.

Consider a universe with $n$ extra spatial dimensions, in
addition to the 3 space and 1 time dimensions we occupy.
A general vacuum solution describing
$n$ such
contracting dimensions and 3 expanding spatial dimensions
was found nearly fourteen years ago \cite{chodos}.
For this solution, the 3 dimensional space begins singular
and decelerates as the universe
evolves.

If the number of extra dimensions exceeds 1,
then there exists
another interesting branch of solutions.
For this other branch, the $n>1$ contracting dimensions drive
the 3 spatial dimensions not only to expand, but what is more,
to inflate.

There is no cosmological constant nor potential driving the inflation.
Thus
higher dimensional theories
exhibit an example of gravity-driven, kinetic inflation \cite{un},
\cite{me1}.  The possibility of kinetic inflation, in the
absence of all conventional potential sources, was
first discussed in the context of dynamical
Planck mass theories.
It is simple to confirm that upon dimensional reduction
the higher dimensional theory is in fact equivalent to
a 4-dimensional theory with a variable Planck mass.

The superstring dilaton also demonstrates
this  property of dynamical
Planck fields.
A kinetic inflation from the dilaton of superstring
theories was discussed independently in Refs. \cite{gv} and
\cite{ven}.
Recently,
there have been interesting studies of
an inflating superstring cosmology in higher dimensions
(see for example \cite{cope}, \cite{forest}).
In those papers it is the dilaton which drives
the inflation.  The extra dimensions
exist only as a consequence of the
nature of string theory.
By contrast, in this paper it is the extra dimensions
themselves which drive the inflation.
As will be discussed,
the extra dimensions act very much like the
superstring dilaton.

While it is possible to drive the three dimensional
world to inflate, a successful completion of kinetic inflation
has yet to be found. For the particular higher dimensional
model discussed in this paper, there will be a graceful
exit problem, as there is for the superstring cosmology \cite{ram}.
For a discussion of the general
obstacle facing the simplest dynamical Planck mass models see
Ref. \cite{me2}.

In the presence of matter sources,
Kaluza-Klein models \cite{big}
are already known to show inflationary behavior.
It was discovered firstly in a radiation dominated
universe
\cite{ra}.  The focus of Refs. \cite{ra} was on the effective
radiation entropy increase.
Even if the net
$N$-dimensional entropy is conserved, the
effective $3$-dimensional   entropy can increase.
In essence, the contracting dimensions
squeeze entropy into the expanding dimensions.
Unfortunately, unless there is some tuning of
parameters, the increase in entropy is
too small to solve the cosmological problems.
Since these original papers,
many source terms have been exploited to fuel inflation
in a universe with extra dimensions.
Anti-symmetric tensor fields predicted from
supergravity \cite{pres}, stringy fluids
\cite{gsv}, as well
as curvatures have all been considered as sources.
\footnote{Since the writing of this paper, I have
discovered reference \cite{ishihara} which considers dust
in a higher-dimensional universe.  The inflationary
solution can be recovered as limiting case of \cite{ishihara}.}

The difference here is that there are no matter sources at all.
The solutions presented in this paper
are strictly vacuum solutions.
There exists only the locally flat spacetime itself.
By stripping the universe down to essentials,
a connection can be made with the dilaton cosmologies
mentioned above.

In addition to the familiar 3 space and 1 time
dimensions, consider the addition of
$n$ extra spacelike dimensions.
Let
the 3-volume expands while the $n$-volume contracts.
It is an initial condition choice to begin
a universe with
all spatial dimensions
expanding, all contracting, or a few of each.
The
naturalness of these initial choices will not be
defended in this paper.
A natural mechanism to compactify additional dimensions
has long been sought in theories which involve higher dimensions.
If such a
mechanism  exists, then the anisotropy of the spacetime
is not a special initial condition but instead is
generated dynamically.  While the mechanics of a realistic
model will likely
be more complicated than those of the simple model studied
here, it seems  fair to conjecture that the basic features
will be generic.

Take the action to be the Einstein theory of gravity
generalized to higher dimensions:
	\begin{equation}
	A=\int d^Nx\sqrt{-g^{(N)}}\left [
	{M_N^2\over 16\pi}{\cal R}^{(N)}
	\right ]
	\ \ ,
	\label{actn}
	\end{equation}
where the total dimension of the spacetime is $N=n+4$.
The $N$-dimensional Planck mass is $M_N$.
The assumption is made that the metric divides simply
into an $N$-dimensional homogeneous but anisotropic metric.
Three spacelike dimensions are taken to expand isotropically
while
the other $n$ spacelike dimensions are taken to contract isotropically.
The metric is of the form
	\begin{equation}
	ds^2_{(N)}=-dt^2+a^2(t)h_{ij}+b^2(t){k}_{mn}
	\ \ .
	\end{equation}
The indices $i,j$ run from $1..3$ and the  indices
$m,n$ run from $4..n+3$.  The scale factor of the $3$-space
is $a(t)$ and that of the $n$-space is $b(t)$.

The vacuum Einstein equations lead to the following set of
equations:
	\begin{eqnarray}
	3{\ddot a \over a}+n{\ddot b \over b}& =& 0
	\label{comp1} \\
	{\ddot a \over a} +(2H_a+nH_b) H_a+{2\kappa^{(3)}\over a^2}&=&0\\
	{\ddot b \over b} +(3H_a+(n-1)H_b) H_b+{(n-1)\kappa^{(n)}\over b^2}
	&=&0
	\ \ ,
	\label{comp}
	\end{eqnarray}
where
	\begin{eqnarray}
	H_a&\equiv& {\dot a\over a} \\
	H_b&\equiv& {\dot b\over b}
	\ \
	\end{eqnarray}
and the local curvatures of the internal
dimensions, $\kappa^{(3)}$ and  $\kappa^{(n)}$, can assume
the values $0,1,-1$.
For the remainder   of this paper all   spatial dimensions
are taken to be locally flat so that
$\kappa^{(3)}=\kappa^{(n)}=0$.
The manifold could still
be closed as is the case with a locally flat toroid.

The set of equations (\ref{comp1})-(\ref{comp}) can be manipulated into
the set of equations,
	\begin{eqnarray}
	H_a^2+nH_aH_b+{n(n-1)\over 6}H_b^2&=&0
	\label{const}\\
	\dot H_a +(3H_a+nH_b)H_a&=&0
	\label{ha}\\
	\dot H_b +(3H_a+nH_b)H_b&=&0
	\label{hb}
	\ \ .
	\label{simp}
	\end{eqnarray}
These equations can be solved very straightforwardly.
It is possible that both $H_a=H_b=0$.
In that case, the universe is empty and flat and
nothing happens.  Alternatively we can assume that neither
$H_a$ nor $H_b$ vanishes.

Firstly, the constraint equation (\ref{const})
can be solved
\footnote{Independently, A. Coley has also found this
to be the most general vacuum solution, though
the inflationary behavior was not
identified \cite{coley}.
}
for $H_b/H_a$,
	\begin{equation}
	{H_b\over H_a}=-\left ( {3n\pm \sqrt{3n^2+6n}
	\over n(n-1) }	\right )
	\label{roots}
	\end{equation}
with $n>1$.  For $n=1$ there is only one solution
to (\ref{const}), $H_b=-H_a$.
If instead $n>1$, there are two roots to this equation.
The lower sign corresponds to the decelerating solution found
14 years ago \cite{chodos} which has since
become a canonical reference (see for instance
\cite{mkkt}, \cite{kolbturner}).
The upper sign on the other hand corresponds to an inflationary
epoch of the 3-space.

It is simple to see that one of the roots corresponds
to an accelerated expansion while the other corresponds
to a decelerated expansion.  From eqn (\ref{roots}) notice that
for
both roots the ratio $H_a/H_b<0$.  Thus if the $n$ extra
dimensions contract, the 3-space expands.
Using (\ref{roots}) in (\ref{ha}) gives
	\begin{equation}
	{\dot H_a\over H_a^2}=\left ({3\pm \sqrt{3n^2+6n}
	\over (n-1)}\right )
	\ \ .
	\label{firstin}
	\end{equation}
If the
upper sign
is operative, eqn (\ref{firstin}) shows that
$\dot H_a>0$.
The Hubble expansion gets ever faster and the scale factor $a$ is
accelerated.
If instead
the
lower sign is operative, then $\dot H_a<0$ for all $n>1$
and the expansion gets ever slower.
Since
$\ddot a/a=\dot H_a+H_a^2$, it follows from (\ref{firstin})
that
	\begin{equation}
	{\ddot a\over a}= \left ({n+2\pm \sqrt{3n^2+6n}
	\over (n-1)}\right )H_a^2
	\ \ .
	\end{equation}
When the lower sign solution applies,
$\ddot a$ is negative and the
scale factor decelerates.

To distinguish the two possibilities, the cosmologies
will be solved separately for the two roots.
For completeness, the decelerating solution
of \cite{chodos} will be reproduced
here.

Integrating (\ref{firstin}) over $dt$ with the lower sign gives
	\begin{equation}
	{-1\over H_a(t)}+{1\over H_a(t_b)}=\left ({3- \sqrt{3n^2+6n}
	\over (n-1)}\right )\Delta t
	\ \ ,
	\label{sol}
	\end{equation}
where $H_a(t_b)$ is a constant of  integration and
$\Delta t=t-t_b$.  For simplicity
$t_b$ will hereafter be taken as $t_b=0$.
{}From eqn (\ref{firstin}) we know that $H_a $ decreases with time
for the lower branch solution.
Consequently,
$1/H_a(0)<1/H_a(t)$ for $t>0$.
An initial condition choice can then be made that
$1/H_a(0)=0$.
With these specifications eqn (\ref{sol}) can be rearranged and integrated
to give
	\begin{equation}
	a=\bar a \left ({t\over \bar t}\right )^r
	\ \ ,
	\label{ad}
	\end{equation}
where $\bar a/\bar t^r$ is a constant of integration and
	\begin{equation}
	r={3+\sqrt{3n^2+6n}\over 3(n+3) }
	\ \ .
	\end{equation}
The exponent $r$ is always less than 1, which
confirms the previous argument that the expansion of the
3-space decelerates.
{}From eqn (\ref{roots}), $b(t)$ is found to be
	\begin{equation}
	b=\bar b\left ({t \over \bar t}\right )^q
	\ \ ,
	\label{bd}
	\end{equation}
where $\bar b$ is a constant of integration and
	\begin{equation}
	q={n-\sqrt{3n^2+6n}\over n(n+3) }
	\ \ .
	\end{equation}

This solution gives a generalized Kasner metric \cite{kasner}
	\begin{equation}
	ds^2=-dt^2+
	\bar a\left ({t\over \bar t}\right )^{2r} dx_{(3)}^2+
	\bar b\left ({t\over \bar t} \right )^{2q} dx_{(n)}^2
	\ \ .
	\end{equation}
The exponents obey the required relations
	\begin{eqnarray}
	3r+nq&=&1\\
	3r^2+nq^2&=&1
	\ \ .
	\end{eqnarray}
This is precisely the cosmology found by Chodos and
Detweiler in 1980 \cite{chodos}.
The decelerating solution is quickly generalized to $d$ isotropically
expanding dimensions and $n$ isotropically contracting dimensions with
	\begin{eqnarray}
	r&\rightarrow &{d+\sqrt{nd(n+d-1)}\over d(n+d) } \\
	q&\rightarrow &{n-\sqrt{nd(n+d-1)}\over n(n+d) }
	\ \ .
	\end{eqnarray}

Now for the accelerating root.  Eqn (\ref{firstin}) can again
be integrated over $dt$, this time for the upper sign solution,
	\begin{equation}
	{-1\over H_a(t)}+{1\over H_a(t_b)}=\left ({3+ \sqrt{3n^2+6n}
	\over (n-1)}\right )\Delta t
	\ \ ,
	\label{sola}
	\end{equation}
where again $H_a(t_b)$ is an integration constant, $\Delta t=t-t_b$,
and $t_b$ is
taken hereafter as zero.  From eqn (\ref{firstin}) it was
argued that the Hubble expansion for the 3 dimensions
grew with time.  As a result, $1/H_a(0)>1/H_a(t)$
for $t>0$.  This constant of integration cannot therefore be
ignored.  Retaining $H_a(0)$, eqn (\ref{sola})
can be solved for $H_a(t)$,
	\begin{equation}
	H_a(t)={H_a(0)\over \left [
	1-t/\bar t\  \right ]}
	\ \ ,
	\label{hat}
	\end{equation}
where the constant $\bar t$ is defined by
	\begin{equation}
	\left (1\over \bar t\right )=\left ({3+ \sqrt{3n^2+6n}
	\over (n-1)} \right ) H_a(0)
	\ \ .
	\end{equation}
\begin{figure}[h]
\vspace{80mm}
\includegraphics{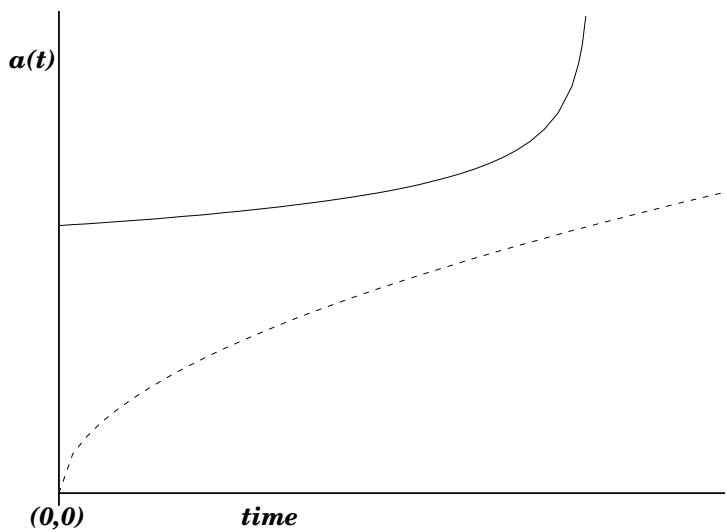}
\vspace{15mm}
\caption{The scale factor $a(t)$ of the 3-space
for both the accelerating solution (solid line)
and the decelerating solution (dashed line).
The time axis is drawn from
$t=0$  to just after $t=\bar t$.
}
\end{figure}

Eqn (\ref{hat}) can be integrated over $dt$ to solve for $a(t)$,
	\begin{equation}
	a={\bar a \over
	\left [1-t/\bar t \ \right ]^{p } }
	\ \ ,
	\label{at}
	\end{equation}
with $\bar a $ a constant of integration and
the exponent $p$ defined by
	\begin{equation}
	p\equiv { -3 + \sqrt{3n^2+6n}\over 3(n+3)}
	\ \ .
	\end{equation}

For $n>1$, $p>0$.
The acceleration can be computed directly from (\ref{at}),
	\begin{equation}
	{\ddot a\over a}=\left ({p\over \bar t \ }\right )
	{p+1\over \bar t}{1\over
	\left [1-t/\bar t \ \right ]^{2 } }\ > 0
	\ \ .
	\end{equation}
This verifies that the 3-volume does grow at an accelerated
rate; that is, it inflates.
Figure 1 sketches the behavior of the scale factor
$a(t)$ with time for both the accelerating solution
of (\ref{at}) and the decelerating solution of (\ref{ad}).

As before, eqn (\ref{roots}) leads to the solution
for $b(t)$
	\begin{equation}
	b(t)=\bar b
	\left [1-t/\bar t \ \right ]^{w}
	\label{bt}
	\end{equation}
where $\bar b$ is a constant of integration
and
	\begin{equation}
	w\equiv {n+
	\sqrt{3n^2+6n}\over n(n+3)}
	\ \ .
	\end{equation}
Figure 2, illustrates the behavior of the
$n$-space scale factor with time for both the accelerating
solution of (\ref{bt}) and the decelerating solution
of (\ref{bd}).

\begin{figure}[h]
\vspace{80mm}
\includegraphics{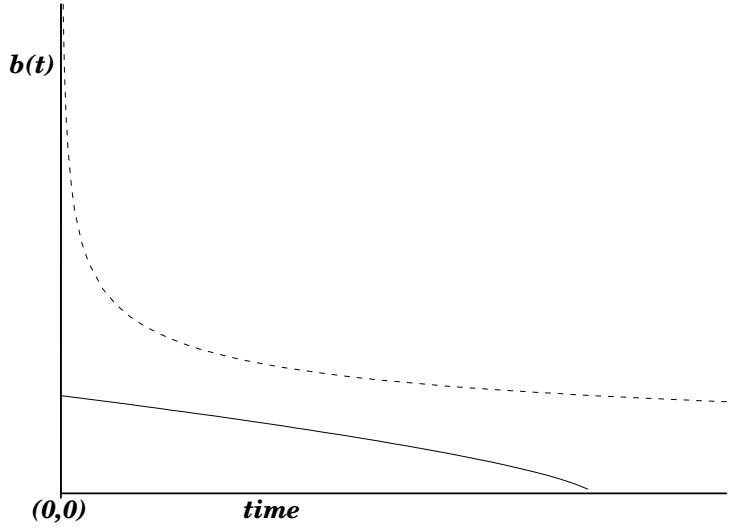}
\vspace{15mm}
\caption{The scale factor $b(t)$ of the $n$-space
for both the accelerating solution (solid line)
and the decelerating solution (dashed line).
The time axis is again drawn from
$t=0$  to just after $t=\bar t$.
}
\end{figure}

This branch gives a generalized Kasner metric of the form,
	\begin{equation}
	ds^2=-dt^2+
	\bar a\left (1-{t\over \bar t}\right )^{-2p} dx_{(3)}^2+
	\bar b\left (1-{t\over \bar t} \right )^{2w} dx_{(n)}^2
	\ \ .
	\label{kasmod}
	\end{equation}
The exponents again obey the required relations
	\begin{eqnarray}
	-3p+n\ w&=&1\\
	3p^2+n\ w^2&=&1
	\ \ .
	\end{eqnarray}
\vfill\eject

The accelerating solution can be simply generalized to $d$ isotropically
expanding dimensions and $n$ isotropically contracting dimensions with
	\begin{eqnarray}
	p&\rightarrow &{-d+\sqrt{nd(n+d -1)}\over d(n+d) } \\
	w&\rightarrow &{n+\sqrt{nd(n+d-1)}\over n(n+d) }
	\ \ .
	\end{eqnarray}

\begin{figure}[h]
\vspace{80mm}
\includegraphics{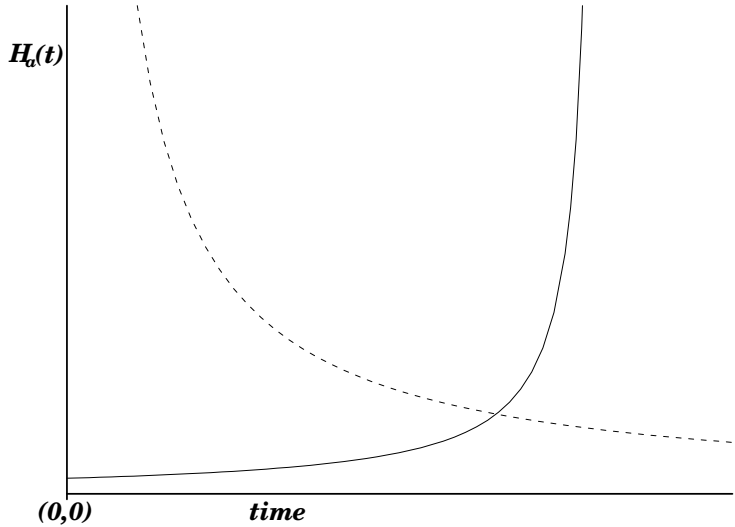}
\vspace{15mm}
\caption{The Hubble expansion of
the $3$-volume, $H_a(t)$
for both the accelerating solution (solid line)
and the decelerating solution (dashed line).
}
\end{figure}

At time $t=t_b=0$, the two scale factors are a finite size,
$a=\bar a$ and $b=\bar b$.  The universe appears to
begin nonsingular
as  $a$ is driven to zero and $b$ is driven to $\infty$ only
in the infinite past, $t=-\infty$.
The expansion of the 3-space gets ever faster as time
moves forward.  As the $n$-space
contracts, the 3-space inflates.
The universe is ultimately ushered into a    future singularity.
At finite time $t=\bar t$, $a\rightarrow \infty$ while
$b\rightarrow 0$.  Also, the Hubble expansions show the singular behavior;
$H_a(\bar t)\rightarrow \infty$ while
$H_b(\bar t)\rightarrow -\infty$ (see figures 3 and 4).
More formally, the spacetime is geodesically incomplete
as geodesics will end in finite affine parameter.

\vfill\eject
\
\begin{figure}[h]
\vspace{80mm}
\includegraphics{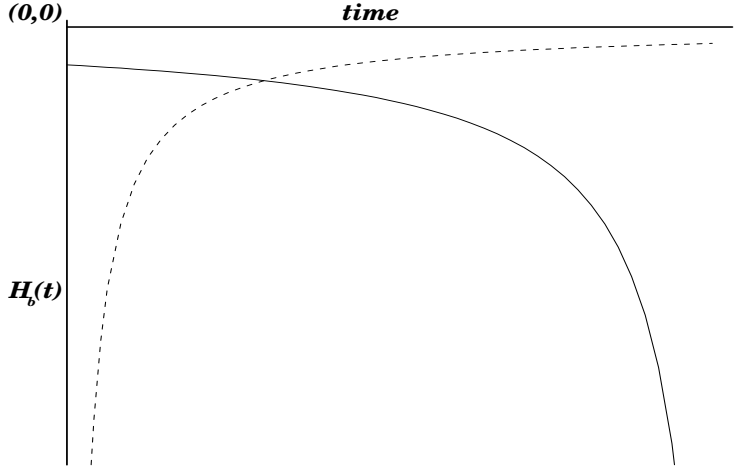}
\vspace{5mm}
\caption{The Hubble contraction of
the $n$-volume, $H_b(t)$
for both the accelerating solution (solid line)
and the decelerating solution (dashed line).
}
\end{figure}

It is well known that upon dimensional reduction, the radius of the
extra dimensions acts just as a dynamical Planck mass
\cite{big}.  Thus
the higher dimensional theory can be related to a generalized
Jordan-Brans-Dicke (JBD) theory \cite{jbd}.
Dynamical Planck fields are often generated
in particle theories, either through simple
quantum corrections or through the full machinery
of theories such as superstrings.
In a higher dimensional universe, the dynamical
Planck mass has a geometric interpretation.
It is related to the radius of the compact internal
dimensions.  To my mind, a geometric source
for the dynamical Planck mass is more
aesthetic than a particle theory source.
It reinforces the suggestion
that the entire scenario is purely and truly
gravity driven.

Begin again with the action for Einstein gravity
generalized to $N=n+4$ dimensions,
	\begin{equation}
	A=\int d^Nx\sqrt{-g^{(N)}}\left [ {M_N^2\over 16\pi}{\cal R}^{(N)}
	\right ]
	\ \ .
	\end{equation}
As is often done,
the action can be  reduced to a four-dimensional theory
by integrating over the extra dimensions,
	\begin{equation}
	S=\int d^4 x\sqrt{-g^{(4)}}
	\int d^nx b^n {M_N^2\over 16\pi}
	\left [ {\cal R}^{(4)}+ n(n-1){\partial_\mu b\partial ^\mu b
	\over b^2} \right ]
	\ \ .
	\label{four}
	\end{equation}
The action (\ref{four}) is equivalent to a 4-dimensional
JBD theory of a dynamical Planck mass,
	\begin{equation}
	\int d^4x \sqrt{-g^{(4)}}
	\left [{\Phi\over 16\pi} {\cal R}^{(4)}-
	\omega(\Phi){\partial_\mu \Phi\partial ^\mu \Phi\over
	16\pi \Phi}
	\right ]
	\ \ ,
	\label{act2}
	\end{equation}
with the definitions
	\begin{equation}
	\Phi\equiv \left (\int d^nx \right )M_N^2 b^n
	\end{equation}
and
	\begin{equation}
	\omega\equiv -1 + {1\over n}
	\ \ .
	\end{equation}
The kinetic coupling parameter
$\omega$ is a negative constant for $n>1$.
Notice, there is no potential nor cosmological constant
in the dimensionally reduced action (\ref{act2}).

The N-dimensional
accelerating solution is thus identical to an example of the
kinetic driven acceleration from a dynamical Planck mass
in ($3+1$)-dimensions.
In references \cite{un} and  \cite{me1} it was shown that
the peculiar kinetic energy of a dynamical Planck mass could drive
the scale factor of the universe to accelerate.
Large families of such theories are possible and can
be identified by a bound on the kinetic couping parameter
$\omega(\Phi)$.  For a vacuum solution, the bound on $\omega(\Phi)$
is
	\begin{equation}
	(1+2\omega /3)^{1/2} \mp 1 -{1\over 3(1+2\omega /3)^{3/2}}
	{\partial \omega(\Phi) \over \partial \ln \Phi}<0
	\ \
	\end{equation}
\cite{me1}.
If $\omega$ is a negative constant, then only for the branch
of solutions corresponding to the upper
sign and a decreasing Planck mass will the universe expand at an
accelerated rate.
In the example of this paper, the
effective $\omega$ is a negative constant,
the branch of solutions corresponds to the required branch,
and the decrease in the Planck mass is
effected by the decrease in the
scale of the extra dimensions.

The
dilaton of superstring theories
in $4$-D
can also provide
an accelerating branch of solutions.  The
superstring dilaton is akin to a JBD theory
with $\omega=-1$.
The dimensionally reduced theory of this paper, giving a JBD theory
with $\omega=-1+1/n$,
is thus very similar
to the superstring dilaton.
Unfortunately, it was found in the string
case that inflation could not
be completed with success \cite{ram} \cite{me2}.
To be specific, our universe today does not connect
smoothly onto the branch of solutions which corresponds
to inflation.  A mechanism to
induce a branch change and provide a graceful
exit seems thus far
to have eluded the string cosmology \cite{ram}.
The graceful exit problem which challenges
superstring inflation will challenge
higher-dimensional inflation  as well.

The graceless exit can be demonstrated simply by
inverting eqn (\ref{roots}) to find $H_a$ for the
inflationary branch of solutions,
	\begin{equation}
	H_a=-{n\over 2}H_b-\sqrt{{(n^2+2n)\over 12}H_b^2
	+{8\pi \over 3M_N^2}\rho}
	\ \ .
	\label{sad}
	\end{equation}
In the above expression, possible contributions from matter
sources have been included in the energy density $\rho$.
Today's universe evolves
roughly as
	\begin{equation}
	H_{a,o}\sim +\sqrt{{8\pi G_o \over 3}\rho_o}
	\ \ .
	\end{equation}
The standard Einstein equations
allow two branches, one expanding and one contracting.
The expanding branch is chosen
as the physically relevant one.
Clearly the inflationary solution of (\ref{sad}) is on
the wrong branch.
Even if a mechanism existed to stabilize the internal dimensions
so that $H_b\rightarrow 0$ and matter domination took over,
it is clear from (\ref{sad}) that
the $3$-volume would ultimately contract, $H_a\rightarrow
-\sqrt{8 \pi\rho/3M_N^2}$.
A branch change is needed if
the inflationary solution
is to connect smoothly onto our expanding cosmology.
\footnote{Even
with source terms driving inflation, as in
Refs. \cite{ra}-\cite{gsv}, such a branch change may be needed
in some cases.}

Regardless of its ultimate fate as
an inflationary model, the higher-dimensional
theory exhibits another interesting feature.
The universe begins cold and empty.  With just
a kick as an initial condition, the cosmology evolves
into a future singularity from a completely regular
state.  This mild
and quiet universe, perhaps a simple thing to create,
evolves into an era of quantum gravity.
Thus a possible
explanation is offered as to why the early universe was
so fiercely energetic.
The cold beginning remains a fascinating
alternative to the standard hot big bang lore.

\bigskip

\centerline{\bf Acknowledgements}

Thank you  to J.R. Bond, A. Coley, N.J. Cornish and
G. Starkman for their thoughts
on this project.
I am also grateful for the additional support of the
Jeffrey L. Bishop Fellowship.

\end{document}